\documentclass[12pt]{iopart}
\usepackage[dvips]{graphicx}

\begin{document}

\title[Waveguide-based OPO source of entangled photon pairs]{Waveguide-based OPO source of entangled photon pairs}

\author{Enrico Pomarico$^{1}$, Bruno Sanguinetti$^{1}$, Nicolas Gisin$^{1}$, Robert Thew$^{1}$, Hugo Zbinden$^{1}$, Gerhard Schreiber$^{2}$, Abu Thomas$^{2}$ and Wolfgang Sohler$^{2}$.}
\address{$^{1}$Group of Applied Physics, University of Geneva, 1211 Geneva,
Switzerland.}
\address{$^{2}$Angewandte Physik, University of Paderborn, 33095
Paderborn, Germany.} \ead{enrico.pomarico@unige.ch}

\begin{abstract}
In this paper we present a compact source of narrow-band
energy-time entangled photon pairs in the telecom regime based on
a Ti-indiffused Periodically Poled Lithium Niobate (PPLN)
waveguide resonator, i.e. a waveguide with end-face dielectric
multi-layer mirrors. This is a monolithic doubly resonant Optical
Parametric Oscillator (OPO) far below threshold, which generates
photon pairs by Spontaneous Parametric Down Conversion (SPDC) at
around 1560\,nm with a 117\,MHz (0.91\,pm)-bandwidth. A coherence
time of 2.7\,ns is estimated by a time correlation measurement and
a high quality of the entangled states is confirmed by a Bell-type
experiment. Since highly coherent energy-time entangled photon
pairs in the telecom regime are suitable for long distance
transmission and manipulation, this source is well suited to the
requirements of quantum communication.
\end{abstract}

\section{Introduction}

Long distance quantum communication is based on the transmission
of entangled photons between remote parties. The transmission
distance is currently limited by photonic absorption in optical
fibre, which increases exponentially with distance. In order to
overcome this limitation, quantum repeater protocols have been
proposed \cite{Briegel98,Simon07}. These protocols consist in the
partition of the communication distance into smaller segments
across which two quantum memories \cite{DeRiedmatten08} can be
efficiently entangled. By performing entanglement-swapping
operations on the intermediate nodes, the entanglement is then
transferred over the entire distance.

In such a scheme, a narrow-band entangled photon pair source is
highly desirable. First of all, it allows the use of atomic
quantum memories, whose spectral linewidths are at the MHz level.
A further advantage of photons with a coherence time larger than
typical detector resolution is that entanglement swapping can be
performed between cw sources, i.e. without the need of
synchronization \cite{Halder07}. Moreover, a high photons'
coherence time increases the tolerance of fibre length
fluctuations and chromatic dispersion.

In order to generate efficient sources of entangled photon pairs,
SPDC in nonlinear crystals is more practical with respect to other
techniques, such as intracavity atomic ensembles \cite{Thompson06}
and bi-exciton cascade emission of quantum dots
\cite{Benson00,Stevenson06}. However, since SPDC sources produce
broad frequency spectra (of the order of THz), the photons'
bandwidth needs to be reduced. Placing a nonlinear crystal inside
an optical cavity, in order to realize OPO-based sources of
narrow-band entangled photon pairs, is a recent approach for
overcoming the limits of SPDC
(\cite{Ou99,Lu00,Goto03,Kuklewicz06,Neergaard07,Bao08,Scholz09b}).
In these systems the bandwidth of the emitted photons is limited
by the cavity and, at the same time, the down conversion process
is enhanced with respect to the single pass scheme
\cite{Ou99,Lu00}.

The OPO sources reported so far produced photon pairs in the
visible regime between 700 and 900\,nm with bandwidths in the MHz
level, comparable to the typical linewidth of atomic quantum
memories. In \cite{Bao08} a narrow band single-mode
polarization-entangled photon source with a linewidth of 9.6\,MHz
at 780\,nm was reported. Only recently \cite{Scholz09b} the first
long-term stable photon pairs OPO source, with a linewidth of
3\,MHz at 894.3\,nm, was made. The schemes reported so far
generate polarization entangled photons in the visible spectrum,
which are not suitable for long distance quantum communication.

In this paper we present a source of narrow-band energy-time
entangled photon pairs at telecom wavelength based on a
Ti-indiffused PPLN waveguide resonator, i.e. a waveguide with
end-face dielectric multi-layer mirrors reflective at telecom
wavelengths.  Our system is the first realization of a monolithic
waveguide-based OPO far below threshold. The integrated optics
approach has significant advantages with respect to bulk OPOs: due
to the high intensity of the pump field with a good overlap to
signal and idler modes even over long distances without any
diffraction, a high non linear efficiency results, which may
surpass that of a bulk source by orders of magnitude. Moreover, no
cavity alignment is necessary and the resonant frequency can be
kept stable by controlling the temperature of the waveguide
\cite{Schreiber01}. In the first part of this paper we describe in
detail the photon source, paying attention to the features of the
frequency spectrum of the generated photons and to the
stabilization of the double resonance condition. Then, we describe
the performance of the photon pair source and characterize the
generated energy-time entangled photon pairs by performing a
Bell-type experiment in the configuration proposed by Franson
\cite{Franson89}. The results are discussed in the conclusion.

\section{The source}
\subsection{The Ti:PPLN waveguide resonator}

\begin{figure}
\begin{center}
\includegraphics[width=0.7\textwidth]{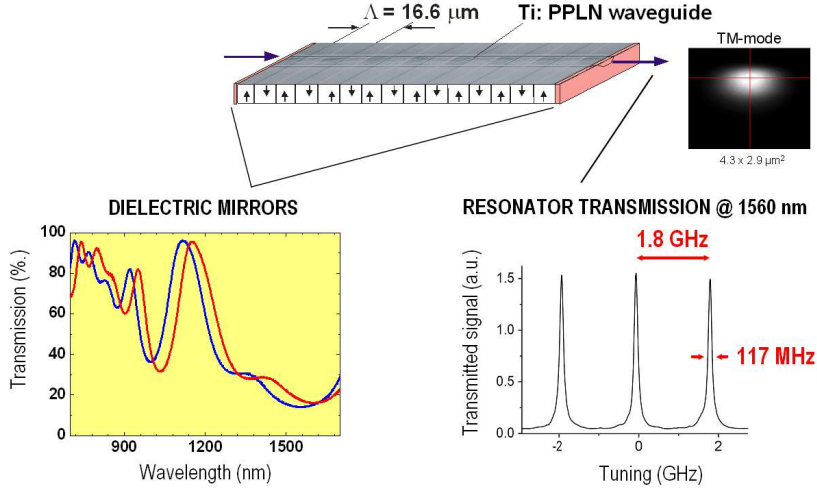}
\caption{\small Features of the Ti:PPLN waveguide resonator. In
the upper part a photograph of the surface of the waveguide and a
photograph of the TM-mode distribution are shown. On the lower
left the transmission spectra of the two mirrors, measured on
LN-substrates simultaneously coated with the waveguide end-faces,
are reported. On the lower right the transmission spectrum of the
waveguide resonator is shown. }\label{fig:waveguide_resonator}
\end{center}
\end{figure}

A 3.6\,cm long waveguide was fabricated by indiffusing (per 9
hours at a temperature of 1060\,$^{\circ}$C) a 6\,$\mu$m wide
vacuum-deposited Ti-stripe of 98\,nm thickness into the -Z- face
of a Z-cut Lithium Niobate (LN) substrate \cite{Schreiber01}.
These parameters were chosen to get a single mode waveguide (at
around 1.55\,$\mu$m of wavelength) of low propagation losses
($\approx$ 0.1\,dB/cm) and a TM-mode distribution (4.3 $\times$
2.9\,$\mu$m$^{2}$, see Figure \ref{fig:waveguide_resonator}) of
good overlap with the mode of standard telecom fibres. The latter
enables low loss fibre coupling and would even allow fibre butt
coupling in the future.

After waveguide fabrication, the sample was periodically poled
with a periodicity of $\Lambda$=16.6\,$\mu$m to get Quasi
Phase-Matching (QPM) for SPDC at a pump wavelength of $\approx$
780\,nm and signal and idler wavelengths at $\approx$
1.55\,$\mu$m.
Figure \ref{fig:waveguide_resonator} shows a photograph of the
surface of such a waveguide; it is slightly etched to make the
domain structure visible. The nearly 1:1 duty cycle of the domain
pattern guarantees a high nonlinear efficiency.

Finally, a stable monolithic waveguide resonator was formed by
depositing dielectric mirrors on the waveguide end-faces. They
consist of alternating SiO$_{2}$ and TiO$_{2}$-layers with
thicknesses defined by a Monte-Carlo optimization to get a high
reflectivity at signal and idler wavelengths (here: $\approx$
85\%), but simultaneously a high transmission for the pump (here:
$\approx$ 80\%). Figure \ref{fig:waveguide_resonator} shows on the
lower left the two mirror characteristics measured on
LN-substrates simultaneously coated with the waveguide end-faces.

The resonator was investigated by measuring its transmission
versus the frequency of a tunable extended cavity laser ($\approx$
1560\,nm). Figure \ref{fig:waveguide_resonator} shows on the lower
right a corresponding result. We observe a Free Spectral Range
(FSR) of 1.8\,GHz (14\,pm) and a Full Width at Half Maximum (FWHM)
of 117\,MHz (0.91\,pm), resulting in a finesse of 15.4. This value
allows to evaluate the waveguide propagation losses as $\alpha =
0.06$\,dB/cm at 1560\,nm. The frequency spectrum of the resonator
can be simply tuned by changing its temperature. In this way its
length and the effective index of refraction are changed leading
to a significant shift of the spectrum. We estimate a wavelength
change of 44.5\,pm/$^{\circ}$C (5.7\,GHz/$^{\circ}$C). In other
words, the shift from one cavity mode to another in the telecom
region implies a change of about 0.3\,$^{\circ}$C at 1560\,nm.

\subsection{Setup}

\begin{figure}
\begin{center}
\includegraphics[width=1\textwidth]{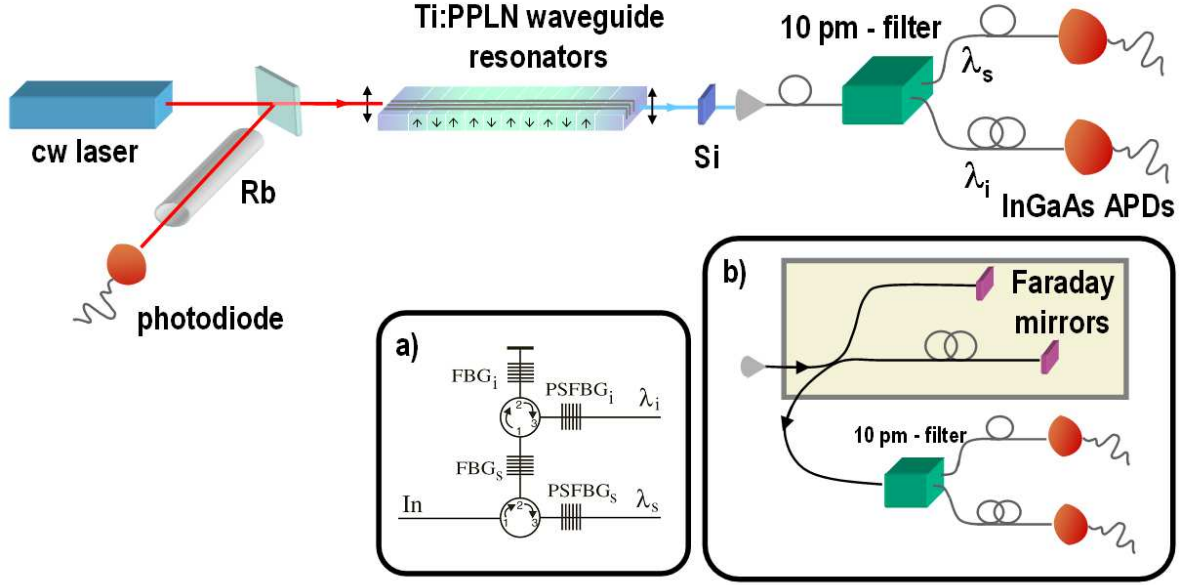}
\caption{\small Setup of the source. A cw mode laser at 780\,nm,
stabilized on a rubidium (Rb) transition, pumps a PPLN waveguide
resonator. Pairs of photons at around 1560\,nm are created by SPDC
and emitted collinearly. Their frequencies correspond to resonant
modes of the cavity. The remaining pump light is blocked by a
silicon filter (Si). The photons, coupled into a single mode
optical fibre, are narrowly filtered (10\,pm - filter) before
being detected by two InGaAs APDs. The filters select the
wavelengths $\lambda_{s}=1559.5$\,nm and $\lambda_{i}=1561.5$\,nm
for the signal and the idler photon respectively. Inset a):
Internal scheme of the 10-pm filter (see text for details). Inset
b): The folded Franson interferometer inserted before the 10-pm
filter for the characterization of the energy-time entangled
states generated by the source.}\label{fig:setup}
\end{center}
\end{figure}

An external-cavity cw diode laser (Toptica DL100), stabilized on a
Rubidium transition (D$_{2}$ of $^{85}$Rb) at
$\lambda_{p}=$780,027\,nm, is used as a pump for producing photon
pairs by SPDC from the PPLN waveguide resonator (Figure
\ref{fig:setup}).

At a temperature of 128.6\,$^{\circ}$C the conditions of QPM and
energy conservation (for which $\omega_{p}=\omega_{s}+\omega_{i}$,
where $\omega_{j}$ (j=p,s,i) refers respectively to the pump,
signal and idler frequency) are fulfilled for the generation of
photon pairs at around 1560 \,nm. Signal and idler photons are
preferentially emitted only if their frequencies also match the
resonant modes of the cavity. Crucial for this goal is the
possibility of properly adjusting the transmission spectrum of the
resonator by changing the temperature of the sample. The
constraint of double resonance affects the frequency spectrum of
the photons emitted by the source. In the next subsection we shall
describe in detail this spectrum and the method for achieving the
double resonance condition at around 1560\,nm.

The pump and the created photons are vertically polarized (type I
eee coupling) and exit the waveguide resonator collinearly. A bulk
high-pass Silicon filter (Si) is placed before coupling into the
fibre in order to block the remaining pump light and has a
transmission of 91\% of the created photons. The signal and the
idler photon are each in a distinct resonant cavity mode. We
distinguish them using a filter (AOS GmbH) which selects two
wavelengths symmetrically around the SPDC degeneracy wavelength
($\lambda_s$=1559.5\,nm and $\lambda_i$=1561.5\,nm) with a
bandwidth of 10\,pm (1.2\,GHz). This value allows us to select
only one cavity mode, as the FSR between them is about 14\,pm
(1.8\,GHz). For the selection of each wavelength (see Figure
\ref{fig:setup}, inset a)) a standard Fibre Bragg Grating (FBG)
reflects the desired wavelength with a bandwidth of $\sim$ 1\,nm
and a circulator directs the reflected light to a Phase-Shifted
Fibre Bragg Grating (PSFBG), which transmits the wavelength with a
10\,pm-bandwidth, as already explained in detail in
\cite{Halder08}. The insertion losses of this filter are 2.8 and
3.8 dB for the two selected wavelengths.

After being filtered, the photons are detected by two InGaAs Avalanche PhotoDiodes (APDs). One APD is in a
free-running mode with 2.1\% quantum efficiency, at which it has
600 dark-counts s$^{-1}$, and 30\,$\mu$s dead time for reducing
after-pulses, whereas the other one operates in gated mode (ID200,
idQuantique) with a quantum efficiency of 7.8\% and a dark-count
probability of 8.0 $\times$ $10^{-6}$ ns$^{-1}$. The detected
signals are sent to a Time to Digital Converter (TDC) for
measuring coincidences between them.

In order to characterize the energy-time entangled states
generated by the source, both signal and idler photons pass
through a single folded Franson interferometer (see Figure
\ref{fig:setup}, inset b)) before being separated by the filter
and detected. The folded Franson interferometer consists of one
fibre coupler (50:50 beam splitter) which splits photons into the
two interferometric arms which have Faraday mirrors at the ends
\cite{Thew02}. Only the signal coming from one output of the
coupler is considered. The interferometer is thermally isolated
and its path difference is $\Delta$L\,$\approx$\,2\,m,
corresponding to a temporal delay of about 10\,ns between the long
and the short arm. This value is smaller than the coherence time
of the pump $\tau_{pump}$ and larger than the coherence time of
the photons $\tau_{coh}$, such that single-photon interference is
negligible.

\subsection{The emitted frequency spectrum and the double resonance
condition}\label{sect:emitted_frequency_spectrum}

\begin{figure}
\begin{center}
\includegraphics[width=0.8\textwidth]{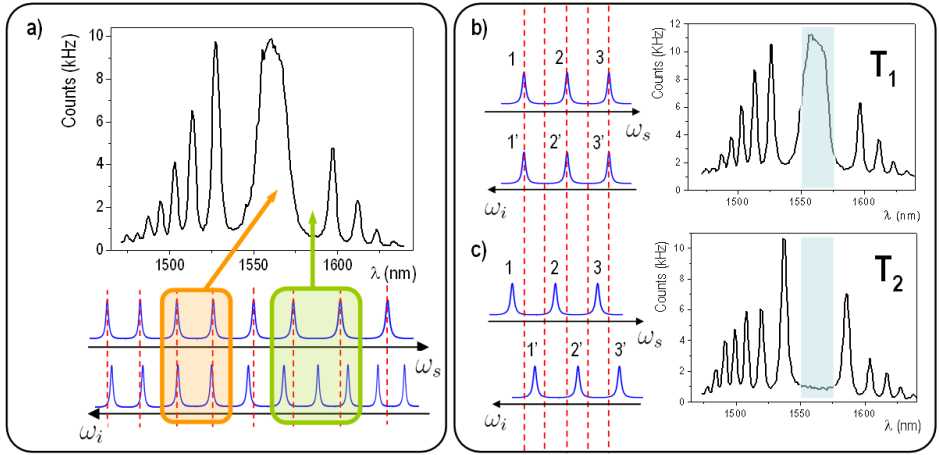}
\caption{\small The emitted frequency spectrum and the double
resonance condition. (a) Above: Envelope of the unfiltered
frequency spectrum of photons produced by SPDC in the waveguide
resonator characterized by cluster emission. Below: Giordmaine and
Miller diagram. The resonant modes for the signal and idler
photons, whose dispersion is exaggerated, are represented along
oppositely oriented axes. When they lie on a vertical line, double
resonance is obtained and a corresponding cluster appears, as for
the central cluster at around 1560 \,nm. (b), (c) Double resonance
condition around 1560 nm. (b) At a certain temperature T$_{1}$ the
modes for the signal and the idler lie on the same vertical lines.
A cluster at around 1560 \,nm appears. (c) By changing the
temperature to T$_{2}$ ($|$T$_{2}$-T$_{1}| \approx$
\,0.07$^{\circ}$C), the signal and idler are no longer aligned
with the vertical lines and the double resonance disappears
together with the central cluster. When the temperature is such
that signal and idler modes move half of the FSR (with about
0.15$^{\circ}$C), double resonance is restored, as in
b).}\label{fig:SPDC_spectrum}
\end{center}
\end{figure}

The parametric radiation has to satisfy four different conditions
(energy conservation, quasi phase-matching and resonance for
signal and idler photons), which affect the frequency spectrum of
the degenerate photons emitted by the waveguide resonators. For an
SPDC source, degenerate photons are spectrally emitted according
to a bell-shape distribution around the frequency $\omega_{p}/2$.
In an OPO system the emission of photons is instead allowed only
within specific spectral intervals. This effect is called cluster
emission \cite{Giordmaine66,Eckardt91,Henderson99}. A typical
envelope of the unfiltered frequency spectrum of emitted photons
is shown in Figure \ref{fig:SPDC_spectrum} (a). Its asymmetry is
essentially due to the decrease of the detection efficiency at
wavelengths above 1600\,nm.

Signal and idler photons can be emitted by the OPO if their
frequencies are symmetric with respect to $\omega_{p}/2$
(condition of energy conservation) and also match two distinct
modes of the resonator. Emission is thus possible only when the
cavity modes for the signal and the idler are symmetric with
respect to $\omega_{p}$/2. However, since the spectral range on
which the photons can be emitted by SPDC is tens of nanometers
wide, it includes thousands of resonant modes. Their relative
spacings can vary considerably within this spectral range because
of dispersion. Thus, cavity modes for signal and idler photons are
symmetric with respect to $\omega_{p}$/2 only in some spectral
regions, called clusters \cite{Eckardt91}. Figure
\ref{fig:SPDC_spectrum} (a) shows this effect for the cluster
around 1560 \,nm. The Giordmaine and Miller diagram
(\cite{Giordmaine66,Eckardt91}) is also shown. The cavity
resonances are plotted for the signal and the idler photon. The
signal frequency increases from left to right, while the idler
goes in the opposite direction. When the transmitted modes of the
resonator for the signal and the idler lie along a vertical line,
energy conservation is satisfied and the corresponding pairs of
photons can be emitted. This is possible only in some intervals,
where a cluster is formed. By changing the temperature of the
sample the sequence of the cavity modes moves and the resonance
condition is automatically lost for certain clusters of
frequencies, but it is obtained for others.

We are interested in the stabilization of the double resonance
condition in the region of frequencies around 1560 nm, at which
photons are filtered. Figure \ref{fig:SPDC_spectrum} (b) and (c)
show the loss of the double resonance in this spectral region by
changing the temperature. In Figure \ref{fig:SPDC_spectrum}(b) a
situation of double resonance at a certain temperature $T_{1}$ is
shown: resonant modes for idler and signal lie on the same
vertical lines. The corresponding frequency spectrum, displaying a
cluster in the region around 1560 \,nm, is shown on the right. If
we change the temperature to $T_{2}$ (such that $|$T$_{2}$-T$_{1}|
\approx$ \,0.07$^{\circ}$C), the modes move to the left for the
signal, and to the right for the idler (Figure
\ref{fig:SPDC_spectrum} (c)). Their superposition on the same
vertical lines is lost and double resonance disappears around 1560
\,nm. If the change of temperature is such that signal and idler
modes move by half their FSR (about 0.15 $^{\circ}$C), the double
resonance is restored and we again observe a cluster around
1560\,nm. In order to maintain the double resonance at 1560\,nm,
unlike previous sources, only an accurate control of the
temperature is required. A thermal stability of better than
$10^{-3}$\,$^{\circ} C$ is guaranteed by performing a P.I.D.
regulation of the temperature using feedback directly from the
signal detected by the free-running detector. Since this is a
side-of-fringe method, we can not lock on the maximum available
signal.

\section{Experimental results}
\subsection{Features of the source} \label{sect:featuressource}
For measuring the coincidences between the photon pairs generated
by the source, the free-running detector provides a start signal
on the TDC and triggers the other detector, which gives the stop
signal. By injecting into the waveguide resonator an effective
pump power of 1.6\,mW (power effectively coupled in the
waveguide), we registered 3400 counts s$^{-1}$ on the first
detector and 5.2 coincidences s$^{-1}$. From these values we
estimate 17 dB of overall losses for the photons directed to the
second detector, in other words only 2\% of the photons generated
inside the resonator reach the second detector. The overall losses
for the photons directed to the first detector are a little
smaller because of the different insertion losses in the 10-pm
filter for the two selected wavelengths. The first source of
photons' losses is inside the waveguide resonator, which we will
try to quantify later on. After being emitted, photons undergo
0.4\,dB of losses because of the passage through the silicon
filter and 5.2\,dB for the coupling into the single mode fibre.
The internal losses of the 10-pm filter are of 2.8 and 3.8\,dB for
the photons directed to the first and the second detector
respectively. We estimated supplementary losses of 2.4\,dB because
of how we lock the temperature of the system. By summing all the
mentioned values, we have 10.8 and 11.8 dB of losses for the
photons directed respectively to the first and the second
detector.

Subtracting, for the second photon, the sum of the above mentioned
losses (11.8\,dB) from the total measured losses (17\,dB), we
estimate that the losses due to the oscillation of the radiation
inside the resonator are 5.2\,dB, that is the 30\% of the
generated photons exit the waveguide resonator. In order to verify
this quantity, we computed the probability that a photon, whose
frequency matches one resonant mode of the cavity, exits the
resonator. By assuming that photons are generated in the middle of
the resonator, the probability of exiting after a certain number
$n$ of round-trips is given by $\sqrt{\alpha L}(1-R)(\alpha L R
)^2$, where $\alpha L$ represents the absorption losses for a
single pass inside the cavity and $R$ the reflectivity of the
mirrors. Therefore, the probability that each photon exits the
cavity can be calculated as:
\begin{equation}
p_{out}=\frac{\sqrt{\alpha L}(1-R)}{1-(\alpha L R)^2},
\end{equation}
which gives a value of 0.43 for R=0.85. This calculation takes in
consideration only photons in perfect resonance with the cavity,
negliging their lorentzian linewidth. This can explain the
discrepancy between the calculated value and the one estimated by
subtracting the overall losses.

Knowing the detection rates and the total losses in the system we
calculate that the source generates 6.6 $\times$ 10$^{6}$ photon
pairs per second in the 10-pm window. The photon coherence time is
2.7\,ns (as we will see in the next paragraph), so the number of
photons per coherence time is 0.02. We keep the pump power low in
order to minimize the emission of double pairs \cite{Scarani05}.
The spectral brightness of the source, calculated considering the
overall losses of the system, corresponds to 17 pairs of photons
emitted per second, per MHz of bandwidth and per mW of effective
pump power (2.2 $\times$ 10$^{3}$ pairs /(s pm mW)) and coupled in
a single mode fibre. Therefore, the signal rate of the source is
not limited by the pump power, but it is highly affected by the
overall losses of the system.

Part of the signal is lost because of the oscillation of the
parametric radiation inside the waveguide resonator. These losses
are related to the time the photons spend in the resonator before
exiting from it. Unfortunately, the resonator bandwidth is related
to this time, therefore we cannot reduce the total losses for a
constant bandwidth. However, the signal rate can be increased
without changing the bandwidth. For example, a single high
reflectivity mirror could be used to reduce the signal which exits
from the wrong side of the resonator. In this case the
reflectivity of the second mirror should be reduced to keep the
finesse, and so the bandwidth constant. Otherwise, one can use a
smaller resonator. In this way the FSR would be increased and, for
keeping the bandwidth constant, the finesse should be increased
properly by improving the mirror reflectivities, allowing to
exploit the cavity enhancement of the signal \cite{Lu00}.

A large part of the signal is also lost in the external optical
components. The integrated optics approach could allow to
drastically reduce these losses, by integrating filters directly
in the waveguide and by fibre butt coupling.

\subsection{Detection of high coherence photons}
\begin{figure}.
\begin{center}
\includegraphics[width=0.7\textwidth]{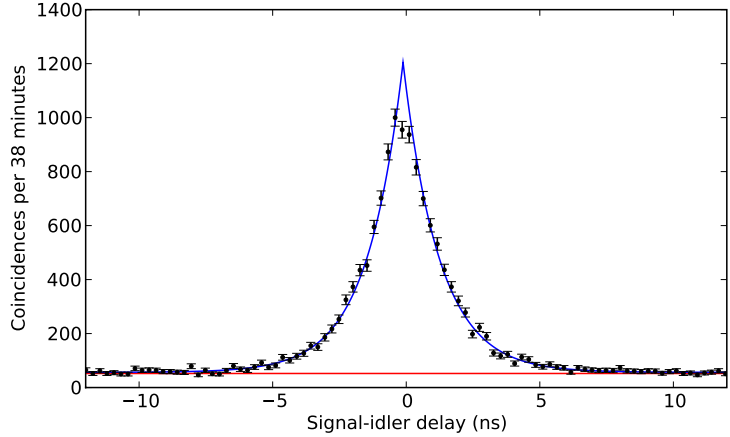}
\caption{\small Coincidences as a function of the temporal delay
between the signal and idler detection. The FWHM of the peak
provides a photons' coherence time of (2.7 $\pm$
0.2)\,ns. Each bin of the histogram is 263 ps-wide. The red line indicates the
level of accidental coincidences.}\label{fig:peak}
\end{center}
\end{figure}
In order to characterize our narrow-band photon pair source, we
performed a time correlation measurement with a TDC. For a
single-mode doubly resonant OPO the cross correlation function can
be expressed, up to multiplicative factors, as $G^{(2)}(\Delta
T)=e^{-2\pi\Delta\nu |\Delta T|}$ \cite{Lu00}, where $\Delta T$ is
the time delay between the idler and signal photons and
$\Delta\nu$ the bandwidth of the down-converted photons. By
fitting the coincidences peak, plotted in Figure \ref{fig:peak},
with the function mentioned above, we obtain
$\Delta\nu=(117\pm7)$\,MHz. Defining the correlation time between
the idler and signal photons as the FWHM of the cross correlation
function, that is $T_{c}=\frac{1.39}{2\pi\Delta\nu}$ \cite{Lu00},
we find $T_{c}=(1.9 \pm 0.1)$\,ns. The width of the cross
correlation function can also be interpreted in terms of coherence
of the down converted photons, which is confirmed by the
interference measurement described in the following section. The
coherence time is $\tau_{coh}=\frac{1}{\pi\Delta\nu}$,
corresponding to $\tau_{coh}=(2.7 \pm 0.2)$\,ns. This value is
almost an order of magnitude higher than previously reported
sources in the telecom regime \cite{Halder08}.

The red line in Figure \ref{fig:peak} indicates the measured level
of the accidental coincidences registered by the TDC, which is in
good agreement with the base-line of the fit. This is due to to
three main factors. One is the contribution due to the noise of
the detectors: accidental coincidences can be registered between
the electronic signals corresponding to the dark-counts of the two
detectors, between a dark-count of one detector and a real
detection on the other detector and viceversa. Accidental
coincidences can also arise because of after-pulses in the first
detector. Another contribution is due to coincidences between
photons of independent pairs. This term is related to the number
of double photon pairs per mode that are generated inside the
crystal. We measured accidental coincidences of 4.7 $\times
10^{-2}$ Hz ns$^{-1}$ due to the noise of the detectors, 0.1
$\times 10^{-2}$ Hz ns$^{-1}$ for after-pulses in the first
detector and we estimated 3.4 $\times 10^{-2}$ Hz ns$^{-1}$
accidental coincidences for independent photon pairs.

\subsection{Characterization of energy-time entangled states}
\begin{figure}
\begin{center}
\includegraphics[width=0.8\textwidth]{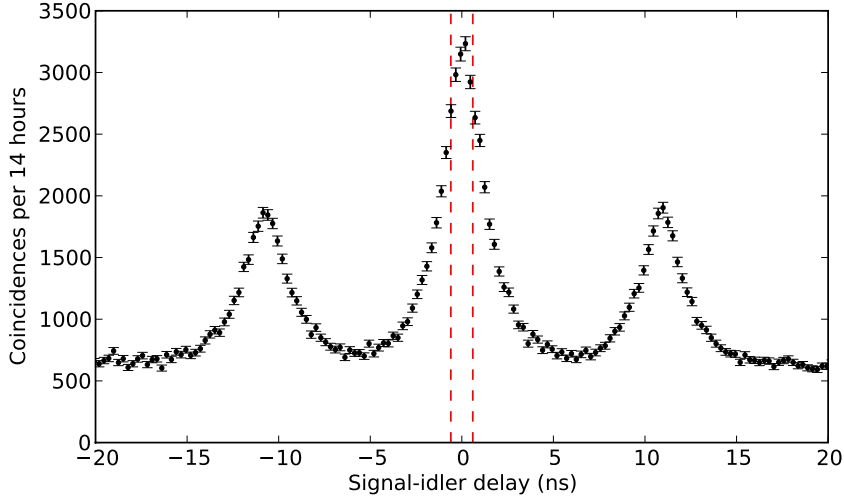}
\caption{\small A histogram of the coincidences as a function of
the signal - idler delay when photon pairs are analyzed by the
folded Franson interferometer. We select coincidences in the
temporal interval corresponding to the central peak, as shown by
the dashed red window, for measuring the interference fringes in
the coincidence count rate. }\label{fig:peaks}
\end{center}
\end{figure}

Since our source is cw pumped, it generates energy-time entangled
photon pairs. Photon pairs may be created at a random time $t$
within the coherence time of the pump, but the uncertainty in the
difference of the emission times of photons of the same pair is
the coherence time of the photons ($\tau_{coh} << \tau_{pump}$).
In other words, the individual energy values of the down-converted
photons are unknown, but their sum is well defined (within the
bandwidth of the pump) by energy conservation.

The energy-time entangled states produced by the source can be
characterized by performing a Bell-type experiment in the
configuration proposed by Franson \cite{Franson89} and measuring
the two-photon interference visibility. In our case, instead of
sending the two photons to two separated analyzer systems based on
equally unbalanced Mach-Zehnder type interferometers, we send
them, as shown in Figure \ref{fig:setup} (inset b)), to only one
interferometer, after which they are filtered and then detected.
This scheme corresponds to a folded Franson interferometer
\cite{Thew02}. Figure \ref{fig:peaks} shows the number of
coincidences as a function of the signal-idler delay. The three
peaks have a temporal separation of approximately 10\,ns, as
expected from the value of $\Delta$L. The events corresponding to
the cases where both the down-converted photons take the short or
the long arm of the interferometer give rise to coincidences in
the central peak. The indistinguishability of these events
provides an interference effect. Side peaks correspond to
coincidences between one photon of the pairs which takes the short
arm while the other takes the long arm, and viceversa.

\begin{figure}.
\begin{center}
\includegraphics[width=0.8\textwidth]{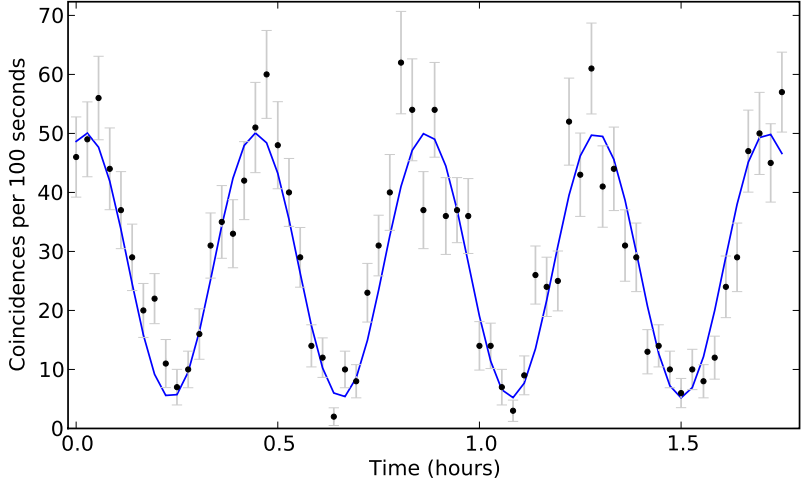}
\caption{\small Coincidences as a function of the time during
which the temperature of the interferometer is scanned. The
measured raw fringe visibility is (81.2$\pm$5.5)\%. The errors
bars are calculated as the standard deviation of the coincidences
values in 100 seconds of integration time for each
point.}\label{fig:fittedfringes}
\end{center}
\end{figure}

By selecting only the coincidences which fall in the temporal
interval corresponding to the central peak, as shown in Figure
\ref{fig:peaks} by the dashed red 1.2\,ns  wide window, and by
scanning the temperature of the interferometer, we measure the
interference fringes in the coincidence count rate (Figure
\ref{fig:fittedfringes}). It is to be noticed that in the folded
Franson interferometer, if we change the temperature we vary the
sum of the phases of both the photons.

The integration time of each point in Figure
\ref{fig:fittedfringes} is 100\,s. The raw fringe visibility is
(81.2$\pm$5.5)\%, sufficient for violating the limit of
$\frac{1}{\sqrt{2}}$ given by Bell inequalities. If we subtract
the accidental coincidences due to the noise of the detectors we
obtain a fringe visibility of (94.4$\pm$5.8)\%. These results
confirm the good quality of the energy-time entangled states
generated by the source.

\section{Conclusions}
The source of energy-time entangled photon pairs presented in this
paper consists of a monolithic doubly resonant OPO system based on
a PPLN waveguide resonator that generates photons by SPDC at
telecom wavelength with a 117\,MHz-bandwidth. We measured a
photons' coherence time of 2.7\,ns and a spectral brightness of 17
emitted photon pairs per second, per MHz of bandwidth and per mW
of effective pump power (2.2 $\times$ 10$^{3}$ pairs /(s pm mW))
coupled in a single mode fibre. The good quality of the
energy-time entangled states we generate is confirmed by the raw
visibility of the interference fringes in the coincidence rate
after a folded Franson interferometer. The signal rate of the
source is highly affected by the overall losses of the system.
Losses due to external optical components could be, instead,
drastically reduce by integrating filters directly in the
waveguide and by fibre butt coupling. Although waveguide-based OPO
technology requires further optimization, this source represents
the first compact and OPO system at telecom wavelength and a first
step towards integrated narrow-band sources of entangled photon
pairs.

\section*{Acknowledgment}
We would like to thank Mikael Afzelius and Hugues de Riedmatten
for useful discussions. This work was supported by the EU projects
QAP and SINPHONIA and by the Swiss NCCR Quantum Photonics.

\section*{References}

\bibliography{enricoBn}
\bibliographystyle{unsrt}

\end{document}